\RequirePackage[T1]{fontenc}
\documentclass[12pt,a4paper]{article} 

\usepackage[height=8.85in,width=6.75in]{geometry}
\usepackage{amsmath}
\usepackage{mathrsfs}
\usepackage{amssymb}
\usepackage{mathtools}
\usepackage{slashed}
\usepackage{braket}
\usepackage{graphicx,rotating}
\usepackage{xcolor}
\usepackage[utf8]{inputenc}
\usepackage{acronym}
\usepackage{bbm}

\definecolor{bleudefrance}{rgb}{0.19, 0.55, 0.91}
\definecolor{desyblue}{HTML}{009EE2}
\definecolor{desyorange}{HTML}{FD8800}
\definecolor{darkred}{rgb}{0.7, 0., 0.}
\definecolor{lightpink}{rgb}{1,0.4,0.4}
\definecolor{lblue}{rgb}{0.384602,0.117763,0.973947}
\usepackage[bookmarksopen,colorlinks=true,
linkcolor=darkred,
citecolor=darkred,
urlcolor=darkred,
linktoc=all]{hyperref}
\newlength{\bibitemsep}
\newlength{\bibparskip}
\let\oldthebibliography\thebibliography
\renewcommand\thebibliography[1]{%
  \oldthebibliography{#1}%
  \setlength{\parskip}{\bibitemsep}%
  \setlength{\itemsep}{\bibparskip}%
}
\usepackage{cite}
\usepackage{times}
\usepackage{courier}
\usepackage{bm,physics}
\usepackage{xcolor}
\usepackage{mdframed}
\usepackage{datetime}
\usepackage{dsfont}
\usepackage{multirow}
\usepackage{cancel}
\usepackage{caption}
\usepackage{subcaption}
\usepackage{graphicx}
\usepackage[compat=1.1.0]{tikz-feynman}
\usepackage[compat=1.1.0]{tikz-feynhand}
\usepackage{tikz}
\usepackage{tikz-cd}
\usepackage{pifont}
\usepackage{romannum}
\AtBeginDocument{\pagenumbering{arabic}}

\renewcommand{\arraystretch}{1.5}

\DeclareMathOperator{\diag}{diag}
\DeclareMathOperator{\udiv}{div}

\newcommand{\PP}[3]{\Phi_{#1}{}^{#2}{}_{#3}}

\newcommand{\AC}[3]{A_{#1}{}^{#2}{}_{#3}}

\newcommand{\FR}[4]{F_{#1}{}_{#2}{}^{#3}{}_{#4}}

\newcommand{\Mpl}{M_\text{Pl}}

\newcommand{\SO}{\mathrm{SO}}

\newcommand{\calJ}{\mathcal{J}}

\newcommand{\calL}{\mathcal{L}}

\newcommand{\calO}{\mathcal{O}}

\newcommand{\calP}{\mathcal{P}}

\newcommand{\trT}[2]{\tr_{(#1#2)}T}

\newcommand{\trdivT}[1]{\tr\udiv_{(#1)}T}

\newcommand{\trQ}[2]{\tr_{(#1#2)}Q}

\newcommand{\trdivQ}[1]{\tr\udiv_{(#1)}Q}

\newcommand{\fT}{\tr\udiv T}

\newcommand{\bae}[1]{\begin{align} #1 \end{align}}

\newcommand{\beae}[1]{\begin{equation}\begin{aligned} #1 \end{aligned}\end{equation}}

\newcommand{\bmte}[1]{\begin{multlined}[t] #1 \end{multlined}}

\definecolor{monza}{HTML}{CF000F}
\definecolor{darkblue}{HTML}{00008b}
\definecolor{darkmagenta}{HTML}{8b008b}

\numberwithin{equation}{section}

\acrodef{EoM}{equation of motion}
\acrodef{CMB}{cosmic microwave background}
\acrodef{EFT}{effective field theory}
\acrodef{SM}{Standard Model}
\acrodef{UV}{ultraviolet}
\acrodef{MAG}{metric-affine gravity}

\begin{document}
\begin{titlepage}

\begin{flushright}
\end{flushright}

\vskip 2.5cm

\begin{center}	
	{\huge \bfseries
        \vskip 1cm
       Towards a classification of UV completable 
        \\
        \vskip .3cm
        Higgs inflation in metric-affine gravity
	}	
	\vskip 2cm
    {\Large
        Yusuke Mikura$^{\ast, }$\footnote{\href{mikura.yusuke.s8@s.mail.nagoya-u.ac.jp}{mikura.yusuke.s8@s.mail.nagoya-u.ac.jp}}
        and Yuichiro Tada$^{\dag,\ast, 
        }$\footnote{\href{tada.yuichiro.y8@f.mail.nagoya-u.ac.jp}{tada.yuichiro.y8@f.mail.nagoya-u.ac.jp}}
    }
    \vskip 1.5cm

    \def\arraystretch{1}
	\begin{tabular}{ll}
        $^\ast$& Department of Physics, 
        Nagoya University, \\ 
        & Furo-cho Chikusa-ku, Nagoya, Aichi 464-8602, Japan\\
        $^\dag$& Institute for Advanced Research,
        Nagoya University, \\ 
        & Furo-cho Chikusa-ku, Nagoya, Aichi 464-8601, Japan
	\end{tabular}
	
    \vskip 1.5cm
	
\end{center}

\noindent
Towards a classification of UV completable Higgs inflation in the framework of parity-even metric-affine gravity, we investigate the particle spectrum of a deformed theory in the large-$N$ limit. In a simple Higgs inflation model in metric-affine gravity, it is known that its UV cutoff is much smaller than the Planck scale. While it calls for UV completion, a concrete example has not yet been found, even with the large-$N$ limit known as a successful technique to complete an original Higgs inflation defined on the Riemannian geometry. This motivates us to study how small deformation of the simple Higgs inflation affects the emergence and properties of dynamical fields particularly in the large-$N$ limit. As a UV theory has to be free of ghosts or tachyons at least around Minkowski space, we perform the parameter search and find the healthy parameter region where a new heavy particle can propagate without these pathologies.

\end{titlepage}

\setcounter{tocdepth}{3}
\tableofcontents

\section{Introduction}

Cosmic inflation is recognized as the modern paradigm of cosmology. While a huge number of theoretical and observational investigations have been made so far, a unanimous understanding of the early universe has not yet been obtained. Among a plethora of models, Higgs inflation~\cite{Bezrukov:2007ep,Barvinsky:2008ia} has been attracting particular attention as this scenario can be realized with the \ac{SM} of particle physics and with a non-minimal coupling between the Higgs boson and gravity:
\bae{\label{eq. Higgs inf}
\calL = \calL_{\mathrm{Higgs}} + \frac{\Mpl^2}{2}R + \xi |H|^2 R ~,
}
where $H$ denotes the Higgs doublet and $R$ is the Ricci scalar.

Higgs inflation is not a unique model but rather a class of models where the Higgs boson triggers the accelerated expansion. Indeed, variants of the original Higgs inflation can be considered by introducing new ingredients in the gravitational sector. The original Higgs inflation, which we call metric-Higgs inflation, adopts the Riemannian geometry as a mathematical foundation of gravity, where only the metric is an independent variable and the connection is given by the Levi-Civita connection, a function of the metric. The algebraic form of the Levi-Civita connection is determined by the torsion-free and metric-compatibility conditions. 
These two conditions are often imposed by hand and therefore can be removed. This brings us a new class of gravitational theories called \ac{MAG}, where both the metric and connection are treated as independent variables. If the connection mass is sufficiently heavy, which is indeed the case in several formulations~\cite {Floreanini:1989hq,Percacci:1990wy}, only the metric part remains as effective degrees of freedom at low energy, making \ac{MAG} consistent with observations.

Even if the connection is (effectively) non-dynamical, the corresponding Euler--Lagrange constraint can change the dynamics of coupled fields, i.e., the Higgs boson. It can
alter the Higgs kinetic and potential terms, making its predictions different from one of the metric-Higgs inflation. Among numerous studies on Higgs inflation in \ac{MAG}, Palatini-Higgs inflation~\cite{Bauer:2008zj} would be a representative example, whose Lagrangian is defined as a simple replacement of the Ricci scalar $R$ in the metric-Higgs inflation~\eqref{eq. Higgs inf} by a curvature scalar $F$ defined by the independent connection $A$:
\bae{\label{eq. Palatini-Higgs inf}
\calL_{\mathrm{PH}} = \calL_{\mathrm{Higgs}} + \frac{\Mpl^2}{2}F + \xi |H|^2 F ~,
}
with 
\bae{\label{eq. curvature F}
\FR{\mu}{\nu}{\rho}{\sigma} \coloneqq \partial_\mu \AC{\nu}{\rho}{\sigma}-\partial_\nu\AC{\mu}{\rho}{\sigma}+\AC{\mu}{\rho}{\lambda}\AC{\nu}{\lambda}{\sigma}-\AC{\nu}{\rho}{\lambda}\AC{\mu}{\lambda}{\sigma} ~.
}
Phenomenology of the metric-Higgs inflation and Palatini-Higgs inflation have been investigated in depth and it is known that their inflationary predictions are in perfect agreement with observations of the \ac{CMB} only with a difference in the prediction of the tensor-to-scalar ratio, given that the coupling between the Higgs boson and gravity is sufficiently strong $\xi \gg 1$ (see, e.g., Ref.~\cite{Tenkanen:2020dge} for the review).

The large coupling is notorious as it results in a very low \ac{UV} cutoff above which perturbative treatment is no longer valid and may induce a violation of perturbative unitarity~\cite{Burgess:2009ea,Barbon:2009ya,Barvinsky:2009ii,Burgess:2010zq,Hertzberg:2010dc,Bezrukov:2010jz,Bauer:2010jg,Lerner:2011it,Kehagias:2013mya,Ren:2014sya,McDonald:2020lpz,Hamada:2020kuy,Antoniadis:2021axu,Mikura:2021clt,Ito:2021ssc,Karananas:2022byw}. 
The low \ac{UV} cutoff is understood as a signal that new physics intervenes around the scale. One possibility of new physics is a weakly coupled \ac{UV} completion, where a problematic low-energy theory is unitarized by the addition of new degrees of freedom whose masses are around the  \ac{UV} cutoff.\footnote{Another possibility is that the model goes into a strongly coupled region and is \ac{UV} completed non-perturbatively.} 
\ac{UV} completion for the metric-Higgs inflation has been well discussed~\cite{Lerner:2010mq,Giudice:2010ka,Barbon:2015fla,Lee:2018esk,He:2018mgb,Ema:2019fdd,Ema:2020zvg}. In particular, Ref.~\cite{Ema:2019fdd} provided an interesting bottom-up approach of \ac{UV} completion with use of the large-$N$ limit, showing that the square of the Ricci curvature, required for one-loop renormalizability~\cite{Salvio:2015kka,Ema:2017rqn,Gorbunov:2018llf}, unitarizes the problematic low-energy theory. The understanding was further deepened in Ref.~\cite{Ema:2020zvg} by viewing the model as a non-linear sigma model with a strongly-curved target space.
The same large-$N$ approach, however, does not give us a desirable result in the case of the Palatini-Higgs inflation as an inclusion of the square of the curvature scalar $F^2$ does not bring new dynamical degrees of freedom and the \ac{UV} cutoff is kept smaller than the Planck scale.\footnote{The large-$N$ limit is applied to Higgs inflation in Einstein--Cartan gravity in Ref.~\cite{He:2023vlj}.}  
Without relying on the large-$N$ limit, one can search for \ac{UV} completion using the fact that the low \ac{UV} cutoff is related to the strong curvature of a target space spanned by the Higgs fields~\cite{Alonso:2015fsp,Nagai:2019tgi,He:2023fko}. It is indeed possible to consider a \ac{UV} theory by adding new scalar fields in a way that the strongly-curved target space is embedded in a higher dimensional flat space~\cite{Giudice:2010ka}. The authors of this paper found that this geometric procedure with one new scalar field does not cure the Palatini-Higgs inflation and its generalization while the metric-Higgs inflation is successfully completed up to the Planck scale~\cite{Mikura:2021clt,He:2022xef}. 
Although it cannot be a rigorous no-go statement on this issue, the absence of a unitarizing field may imply that the Palatini-Higgs inflation does not come out as a perturbative low-energy theory. 

In the circumstances, it is of great importance to investigate a new Higgs inflation model that can be safely completed up to the Planck scale in the realm of \ac{MAG}.
In a UV theory, the connection would be promoted to a dynamical degree of freedom and is expected to uplift the UV cutoff. However, prior to the study of the new UV cutoff, one has to notice that some components of the connection sometimes propagate pathologically being ghosts or tachyons due to the nature of the Lorentz group. It is therefore reasonable to study in what parameter space the new Higgs inflation is connected to a ghost- and tachyon-free \ac{MAG} around Minkowski space. Non-pathological propagation of the connection is a necessary condition to have a well-behaved UV theory and the parameter search may help us to understand the peculiarity of the high-energy aspect of the Palatini-Higgs inflation.
In this work, we focus on a small deformation of the Palatini-Higgs inflation, where the word ``small deformation'' means that a low-energy theory is defined by using the same operators as the Palatini-Higgs inflation while their coefficients are unfixed, and consider its \ac{UV} theory by taking quantum corrections into account in the large-$N$ limit.\footnote{While one may be interested in the inflationary phenomenology in the ``deformed" theory~\cite{Rigouzzo:2022yan,Zell}, we would like to first focus on whether the theory will be healthy or not.} By exploring the particle spectrum of the \ac{UV} theory, we study in what parameter region \ac{UV} particles are free of pathological propagation.

The paper is organized as follows. In Sec.~\ref{sec. Neighboring theory}, we define a small deformation of the Palatini-Higgs inflation and introduce a \ac{UV} theory by the addition of new operators induced by quantum corrections. Then, in Sec.~\ref{sec. SPO}, we present the basics of the spin projector formalism which is a powerful tool to study the particle spectrum, and provide conditions for particles to be ghost- and tachyon-free around Minkowski space. In Sec.~\ref{sec. particle spectrum}, we turn to examples with kinematical constraints for simplicity. Finally, we conclude in Sec.~\ref{sec. Conclusions}.
Throughout the paper, we adopt the natural unit $c=\hbar=1$ and $\eta_{\mu\nu} = \diag(-1,1,1,1)$ is used for the sign of the Minkowski metric.

\section{Small deformation of Palatini-Higgs inflation}\label{sec. Neighboring theory}
As mentioned in the introduction, the Palatini-Higgs inflation is described by the curvature scalar $F$ defined by Eq.~\eqref{eq. curvature F}.
To define a deformed theory, let us first introduce the distortion tensor $\Phi$ by
\bae{
	\Phi_{\mu}{}^{\rho}{}_{\nu} \coloneqq A_{\mu}{}^{\rho}{}_{\nu}-\Gamma_{\mu}{}^{\rho}{}_{\nu} ~,
}
with which the curvature scalar is rewritten as
\bae{
	F= R
	+g^{\mu\nu}\left(
	\nabla_\rho\PP{\mu}{\rho}{\nu}
	-\nabla_\mu\PP{\lambda}{\lambda}{\nu}
	+\PP{\rho}{\rho}{\tau}\PP{\mu}{\tau}{\nu}
	-\PP{\mu}{\rho}{\lambda}\PP{\rho}{\lambda}{\nu}
	\right) ~,
}
where $\nabla$ is the covariant derivative formed with the Levi-Civita connection:
\bae{
\nabla \coloneqq \partial + \Gamma ~.
}
In terms of the distortion tensor $\Phi$, \ac{MAG} can be viewed as metric theories with a three-index matter field living on the Riemannian manifold. It is often the case that we impose a constraint where the distortion tensor is either antisymmetric or symmetric. The constraint is partially for simplicity but, in a certain class of theories, one can utilize a symmetry of a theory to remove either an antsymmetric or a symmetric part of the distortion tensor without loss of generality. In the case of the Palatini-Higgs inflation, the projective symmetry, a shift of a generic connection, plays this role. In this case, it is more convenient to define the torsion $T$ and non-metricity $Q$ by
\bae{
    T_{\alpha\beta\gamma}  \coloneqq \Phi_{\alpha\beta\gamma} - \Phi_{\gamma\beta\alpha} ~, \quad
    Q_{\alpha\beta\gamma}  \coloneqq  \Phi_{\alpha\beta\gamma} + \Phi_{\alpha\gamma\beta} ~.
}
Note that the torsion is antisymmetric under the exchange of the first and third indices while the non-metricity is symmetric in the last two indices. 
In this paper, we call theories without the non-metricity \emph{antisymmetric \ac{MAG}} and theories without the torsion \emph{symmetric \ac{MAG}}.

\subsection{In antisymmetric metric-affine gravity}
Let us define a deformed theory in antisymmetric \ac{MAG} by imposing that the symmetric part of the distortion tensor vanishes. 
In terms of the torsion, the curvature scalar can be expanded as
\bae{\label{eq. curvature torsion}
F = R + 2 \trdivT1 + \frac{1}{4}M^{TT}_1 + \frac{1}{2}M^{TT}_2 - M^{TT}_3 ~,
}
where $\udiv_{(i)}T_{ab}$ is the divergence of $T$ on the $i$-th index, ``$\tr$'' is the trace over the given indices, e.g., $\trT12_\mu=T_\nu{}^\nu{}_\mu$, and $M^{TT}$ are mass terms defined by
\bae{
M^{TT}_1 \coloneqq T_{\mu\nu\rho}T^{\mu\nu\rho} ~, \quad 
M^{TT}_2 \coloneqq T_{\mu\nu\rho}T^{\mu\rho\nu} ~, \quad 
M^{TT}_3 \coloneqq \trT12_\mu \trT12^\mu ~.
}
Motivated by the form of Eq.~\eqref{eq. curvature torsion}, we deform the Palatini-Higgs inflation to
\bae{\label{Eq. IR theory torsion}
\calL_{{\mathrm{IR}}}^{\mathrm{T}} \coloneqq \frac{\Mpl^2}{2}\left(1+\xi \frac{\phi^2}{\Mpl^2}\right)\left(R + x \trdivT1 -c_1^{TT} M^{TT}_1 - c_2^{TT} M^{TT}_2 - c_3^{TT} M^{TT}_3 \right) ~,
}
where $\phi_I$ is a $N$-component field 
with $\phi^2 \coloneqq \delta^{IJ}\phi_I\phi_J$ 
($N=4$ corresponds to the SM Higgs), $x$ and $c_i^{TT}$ $(i=1, 2, 3)$ are unfixed parameters, and $\xi$ is the non-minimal coupling which is assumed to be large $\xi \gg 1$ as it is required in many cases to satisfy the \ac{CMB} constraint with a moderately large self-coupling. 
Note that the Palatini-Higgs inflation can be recovered by taking a specific choice of parameters:
\bae{\label{eq. Palatini-Higgs limit torsion}
    x=2 ~, \quad 
    \pqty{ c_1^{TT}, c_2^{TT}, c_3^{TT} }
    = \pqty{ -\frac{1}{4}, -\frac{1}{2}, 1 } ~.
}
Since the action~\eqref{Eq. IR theory torsion} contains at most a first order derivative of $T$ in time, the torsion is a constraint field at a classical level.

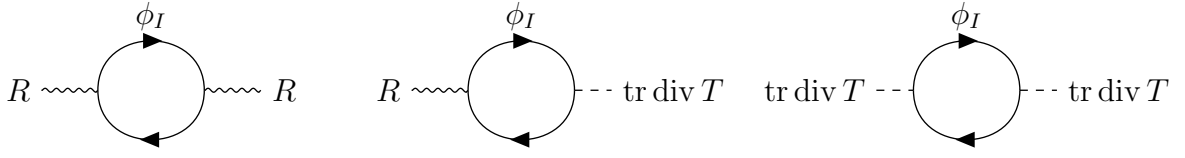
\begin{figure}
	\centering
	\begin{minipage}
 {0.3\linewidth}
	\centering
	\begin{tikzpicture}[scale=1]
	  \begin{feynhand}
        \vertex (x1) at (-1.75,0){$R$};
		\vertex[crossdot] (p) at (-0.7,0);
		\vertex (q) at (0.7,0);
		\vertex (w1) at (1.75,0) {$R$};
  		\propag [photon] (x1) to (p);
		\propag [fer] (p) to [in=90, out=90, looseness=1.6, edge label=$\phi_I$] (q);
  		\propag [antfer] (p) to [in=-90, out=-90, looseness=1.6] (q);
		\propag [photon] (q) to (w1);
	  \end{feynhand}
	\end{tikzpicture}
  \end{minipage}
  \begin{minipage}
  {0.3\linewidth}
	\centering
	\begin{tikzpicture}[scale=1]
	  \begin{feynhand}
	\vertex (x1) at (-1.75,0){$R$};
		\vertex[crossdot] (p) at (-0.7,0);
		\vertex (q) at (0.7,0);
  	    \vertex (w1) at (2,0) {$\fT$};
  		\propag [photon] (x1) to (p);
		\propag [fer] (p) to [in=90, out=90, looseness=1.6, edge label=$\phi_I$] (q);
  		\propag [antfer] (p) to [in=-90, out=-90, looseness=1.6] (q);
		\propag [scalar] (q) to (w1);
	  \end{feynhand}
	\end{tikzpicture}
  \end{minipage}
  \begin{minipage}
  {0.3\linewidth}
	\centering
	\begin{tikzpicture}[scale=1]
	  \begin{feynhand}
	\vertex (x1) at (-2,0){$\fT$};
		\vertex (q) at (0.7,0);
  	    \vertex (w1) at (2,0) {$\fT$};
  		\propag [scalar] (x1) to (p);
		\propag [fer] (p) to [in=90, out=90, looseness=1.6, edge label=$\phi_I$] (q);
  		\propag [antfer] (p) to [in=-90, out=-90, looseness=1.6] (q);
		\propag [scalar] (q) to (w1);
	  \end{feynhand} 
	\end{tikzpicture}
  \end{minipage}
  \caption{Diagrams to be regulated with counterterms. The left one requires the inclusion of the $R^2$ operator. The middle one corresponds to the kinetic mixing between the spin-0 component of the graviton and the torsion. The last one is related to the kinetic term of the torsion. The arrow denotes the flow of a flavor index $I$ of the Higgs.}
  \label{fig:divergent diagrams}
  \end{figure}

Quantum corrections can modify the classical action~\eqref{Eq. IR theory torsion} by generating the kinetic terms of the torsion and mixing terms with gravity. Since the number of propagating degrees of freedom can change, the addition of the quantum effects can be understood as \ac{UV} completion if the new fields uplift the low cutoff. With the IR Lagrangian~\eqref{Eq. IR theory torsion}, a UV lagrangian quadratic in curvature and torsion is expected to take the following form:
\beae{\label{eq. UV lag torsion}
    \calL_{\mathrm{UV}}^{\mathrm{T}} \coloneqq
    \bmte{\frac{\Mpl^2}{2}R + \alpha R^2 -\frac{1}{2}b^{RT}_5 R \trdivT1 -\frac{1}{2}b^{TT}_9 (\trdivT1)^2 
    \\
    -\frac{\Mpl^2}{2}\left(m_1^{TT} M^{TT}_1 + m_2^{TT} M^{TT}_2 + m_3^{TT} M^{TT}_3 \right) ~,}
}
where $\pqty{\alpha, b^{TT}_9, b^{RT}_5, m^{TT}_i}$ are new unknown parameters.\footnote{The subscript numbering follows the convention used in Ref.~\cite{Baldazzi:2021kaf}.
}
We omit the Higgs sector since the Higgs fields appear as a bilinear form and are therefore decoupled from the torsion in the quadratic action. Here two remarks are in order. First, we neglect another quadratic gravitational operator $R_{\mu\nu}R^{\mu\nu}$ because this operator becomes
relevant only around the Planck scale even with a large $\xi$ as explained in Ref.~\cite{Ema:2020zvg}. Second, the \ac{UV} parameters $\pqty{\alpha, b^{TT}_9, b^{RT}_5, m^{TT}_i}$ are related to each other and to the parameters in the low-energy theory in a complicated way.
If we rely on the large-$N$ limit, diagrams with the Higgs running in a loop (Fig.~\ref{fig:divergent diagrams}) contribute dominantly.
In this case, we would have a relatively simple relation among the \ac{UV} parameters and the parameters in the deformed theory, which can be seen from
\bae{\label{eq. UV lag largeN}
\calL_{\mathrm{UV}, N}^{\mathrm{T}} \coloneqq \calL_{{\mathrm{IR}}}^{\mathrm{T}} + \alpha \left(R + x \trdivT1 -c_1^{TT} M^{TT}_1 - c_2^{TT} M^{TT}_2 - c_3^{TT} M^{TT}_3  \right)^2 ~,
}
where the value of $\alpha$ is of order $\xi^2$.

\subsection{In symmetric metric-affine gravity}
In the same way, we can define a deformation in symmetric \ac{MAG}. With the torsion-free condition, the curvature scalar $F$ can be expressed in terms of the non-metricity $Q$ as 
\bae{\label{eq. curvature Q}
F = R - \trdivQ1 + \trdivQ2 + \frac{1}{4}M^{QQ}_1 - \frac{1}{2}M^{QQ}_2 -\frac{1}{4} M^{QQ}_3 + \frac{1}{2} M^{QQ}_5 ~,
}
where we introduce the shorthand notation
\bae{
\trdivQ{1} \coloneqq g^{\mu\nu}\nabla_\rho Q^{\rho}{}_{\mu}{}_{\nu} ~, \quad 
\trdivQ{2} \coloneqq g^{\mu\nu}\nabla_\rho Q_{\mu}{}^{\rho}{}_{\nu} ~,
}
and $M^{QQ}$ are the mass terms of the non-metricity defined by\footnote{There exists another mass term labelled by $i=4$ in \ac{EFT} of \ac{MAG}~\cite{Baldazzi:2021kaf}. However, it does not appear in the decomposition of the curvature scalar $F$.
}
\beae{
M^{QQ}_1 &\coloneqq Q^{\rho\mu\nu} Q_{\rho\mu\nu} ~,&
\quad
M^{QQ}_2 &\coloneqq Q^{\rho\mu\nu} Q_{\nu\mu\rho} ~,
\\
M^{QQ}_3 &\coloneqq \trQ23^{\mu} \trQ23_\mu ~, &
\quad
M^{QQ}_5 &\coloneqq \trQ23^{\mu} \trQ12_\mu ~.
}
Here $\tr_{(ij)}Q^a$ denotes the trace of the non-metricity on the $i$-th and $j$-th index. With Eq.~\eqref{eq. curvature Q}, we define a deformed theory of the Palatini-Higgs inflation in symmetric \ac{MAG} by 
\beae{\label{Eq. IR theory non-metricity}
\calL_{{\mathrm{IR}}}^{\mathrm{Q}} \coloneqq \frac{\Mpl^2}{2}\left(1+\xi \frac{\phi^2}{\Mpl^2}\right)\left(R + y \trdivQ{1} + z \trdivQ{2} -\sum_{i} c_i^{QQ} M^{QQ}_i \right) ~, \quad (i=1,2,3,5)
}
where $\pqty{y,z, c^{QQ}_i}$ are unfixed parameters.
We note that the Palatini-Higgs inflation is recovered if one takes
\bae{\label{eq. Palatini-Higgs limit nonmetricity}
    y=-1 ~, \quad 
    z=1 ~, \quad 
    \pqty{
    c_1^{QQ}, c_2^{QQ}, c_3^{QQ}, c_5^{QQ}
    } = 
    \pqty{-\frac{1}{4}, -\frac{1}{2}, -\frac{1}{4}, \frac{1}{2}
    }
~.
}

As in the torsion case, quantum corrections are expected to generate the kinetic terms of the non-metricity. Recalling that the non-metricity has two different ``$\tr\udiv$'' terms, we expect that a \ac{UV} Lagrangian takes the following form:
\beae{\label{eq. UV lag nonmetricity}
    \calL_{\mathrm{UV}}^{\mathrm{Q}} \coloneqq  
    \bmte{ \frac{\Mpl^2}{2}R + \alpha R^2 -\frac{1}{2}b^{RQ}_6 R \trdivQ1 -\frac{1}{2}b^{RQ}_7 R \trdivQ2
    -\frac{1}{2}b^{QQ}_{14} (\trdivQ1)^2 \\
    -\frac{1}{2}b^{QQ}_{15} (\trdivQ2)^2 -\frac{1}{2}b^{QQ}_{16} \trdivQ1\cdot \trdivQ2
    -\sum_{i}\frac{\Mpl^2}{2}m_i^{QQ} M^{QQ}_i ~, }
}
where $m^{QQ}_i$ with $(i=1,2,3,5)$ are new mass parameters.
To the leading order in the large-$N$ limit, the quantum corrections can be introduced in a compact way as
\bae{\label{eq. large N nonmetricity}
\calL_{\mathrm{UV}, N}^{Q} \coloneqq \calL_{{\mathrm{IR}}}^{\mathrm{Q}} + \alpha \left(R + y \trdivQ{1} + z \trdivQ{2} -\sum_{i}c_i^{QQ} M^{QQ}_i \right)^2 ~,
}
where $\alpha$ is a parameter of order $\xi^2$.

\section{Spin projector formalism and non-pathological propagation}\label{sec. SPO}
The torsion and non-metricity are reducible under the rotation group $\SO(3)$, meaning that they include several particles. In a generic \ac{EFT} setup, some of the particles propagate pathologically being ghosts or tachyons due to the non-definiteness of the Minkowski metric. While ghosts and tachyons are problematic to have a unitary theory, they can be removed from the spectrum by properly tuning coefficients of the \ac{EFT}. A powerful tool to analyze the spectrum of MAG around the Minkowski spacetime is the spin projector formalism. It enables us to decompose multi-index tensors and the kinetic structure of a theory into irreducible representations of the group $\SO(3)$.\footnote{There exist several works that utilize the spin projector formalism to search for non-pathological \ac{MAG}~\cite{Percacci:2020ddy,Lin:2020phk,Marzo:2021esg,Marzo:2021iok,Mikura:2023ruz,Mikura:2024mji,Barker:2024ydb}. See also Ref.~\cite{Barker:2024juc} for the computing software.}
The spin projectors for two-index tensors were introduced in Refs.~\cite{Rivers:1964nfl,barnes,Aurilia:1969bg}, for antisymmetric three-index tensors in Ref.~\cite{Sezgin:1979zf}, and further generalized to arbitrary three-index tensors in Ref.~\cite{Percacci:2020ddy}.

\subsection{Spin projector formalism}\label{sec SPO formalism}
It is known that each representation of $\SO(3)$ is labeled by an integer spin $\mathcal{J} = 0,1,2,\ldots$, and parity $\calP =\pm$. 
Let $\varphi_A$ be a multi-index tensor where the subscript $A$ contains multiple Lorentz indices, e.g., two indices for a two-index tensor and three indices for a three-index tensor. The multi-index tensor can be decomposed into small pieces by spin projectors $P_{ii}(\calJ^\calP)$ through
\bae{\label{Eq. completeness tensor}
    \varphi_A =\sum_{i, \calJ,\calP} P_{ii}(\calJ^\calP){}_A{}^B \varphi_B ~,
}
where $\calJ$ and $\calP$ denote spin and parity, and the index $i$ labels different components in the same spin-parity sector.
It is clear from Eq.~\eqref{Eq. completeness tensor} that the spin projectors satisfy the completeness relation
\bae{\label{Eq. completeness}
\sum_{i, \calJ,\calP} P_{ii}(\calJ^\calP)= \mathbb{I} ~,
}
where $\mathbb{I}$ is the identity operator and we omit the Lorentz indices for brevity. In addition to the spin projectors that single out each representation, we can define intertwiners $P_{ij}(\calJ^\calP)$ that implement the mapping between representations with the same spin and parity but different $\pqty{i,j}$ indices. We collectively refer to all the projectors and intertwiners as the spin projectors.

The spin projectors form complete bases. On top of the above completeness relation, they satisfy the orthonormality relation
\bae{\label{Eq. orthonormality}
P_{ij}(\calJ^\calP){}_A{}^B P_{kl}(\calJ^{\prime\, \calP^\prime}) {}_B{}^C=\delta_{\calJ\calJ^\prime}\delta_{\calP \calP^\prime}\delta_{jk}P_{il}(\calJ^\calP){}_A{}^C ~,
}
and the hermiticity relation 
\bae{
P_{ij}(\calJ^\calP){}_A{}^B = \left(P_{ji}(\calJ^\calP){}^B{}_A\right)^\ast ~.
}
As the orthonormality relation is an equation of the form
\bae{
    P^2 = P ~,
}
eigenvalues of the spin projectors are $\lambda=0$ and $\lambda=1$. Therefore the value of the trace of the spin-$\calJ$ projector is identical to the dimension of representation $d(\calJ)$ as
\bae{\label{Eq. Trace of projectors}
    \Tr P(\calJ^\calP){}_A{}^B = d(\calJ)= 2\calJ+1 ~.
}

Higher rank tensors can be constructed by the product of vectors. Thus the spin projectors can be constructed with the longitudinal and transverse projectors on vectors, defined by
\bae{\label{def of L and T}
    L_\mu{}^\nu = \frac{q_\mu q^\nu}{q^2}~, \quad  T_\mu{}^\nu =\delta_\mu^\nu - L_\mu{}^\nu ~.
	}
Table~\ref{Table irreps} lists irreducible representations for the three-index and symmetric two-index tensors and their corresponding projectors. See Ref.~\cite{Percacci:2020ddy} for explicit forms of the spin projectors. 
\begin{table}
    \begin{center}
    \renewcommand{\arraystretch}{1.2}
    \begin{tabular}{|c|c|c|c|c|}
    \hline
    & $ts$ & $hs$ & $ha$ & $ta$ \\
    \hline
    $TTT$ & $3^-$, $1^-_1$ & $2^-_1$, $1^-_2$ & $2^-_2$, $1^-_3$ & $0^-$ \\
    \hline
    $TTL+TLT+LTT$ & $2^+_1$, $0^+_1$   & -  & -  & $1^+_3$  \\
    \hline
    $\frac{3}{2} LTT$ & - & $2^+_2$, $0^+_2$ & $1^+_2$, & - \\
    \hline
    $TTL+TLT- \frac12 LTT$ & - & $1^+_1$  & $2^+_3$, $0^+_3$  & -  \\
    \hline
    $TLL+LTL+LLT$    & $1^-_4$   &  $1^-_5$   & $1^-_6$   &  -  \\
    \hline
    $LLL$   & $0^+_4$   &  -   & - &  -  \\
    \hline
    \end{tabular}
    \quad
    \begin{tabular}{|c|c|}
    \hline
     & $s$ \\
    \hline
    $TT$ & $2^+_4$, $0^+_5$ \\
    \hline
    $TL$  & $1^-_7$  \\
    \hline
    $LL$   &  $0^+_6$  \\
    \hline
    \end{tabular}
    \end{center}
    \caption{$\SO(3)$ spin content of projection operators for a generic three-index tensor and symmetric two-index tensor ($ts$ = totally symmetric, $hs$ = hook symmetric, $ta$ = totally antisymmetric, $ha$ = hook antisymmetric). The symbols $L$ and $T$ in the first column refer to the longitudinal and transverse projectors for vectors, defined by Eq.~\eqref{def of L and T}.
    The subscripts distinguish different instances of the same representation. The non-consecutive numbering follows from the conventions of Ref.~\cite{Baldazzi:2021kaf}.}
    \label{Table irreps}
\end{table}
\subsection{Conditions without pathology}
For the analysis of the particle spectrum, let us expand the metric around the Minkowski space as
\bae{
g_{\mu\nu} \simeq \eta_{\mu\nu} +\frac{2}{\Mpl}h_{\mu\nu} ~,
}
and consider a theory that consists of terms quadratic in the curvature and three-index tensor. In the Fourier space, the quadratic action takes the form
\bae{\label{eq. quadratic O}
S^{(2)} = \frac{1}{2}\int \frac{\dd[4]{q}}{(2\pi)^4}
~ X^T \cdot \calO \cdot X ~,
}
where $X=(\Phi_{\mu\nu\rho}, h_{\mu\nu})$ and $\calO$ is constructed only with the flat metric $\eta_{\mu\nu}$ and with momentum $q_\mu$. Note that the kinetic operator $\calO$ is a $2\times 2$ matrix.

Expanding the kinetic operator $\calO$ by the spin projectors, one can rewrite the quadratic action as
\bae{\label{Eq. decomposed quadratic form}
S^{(2)}= \frac12\int \frac{\dd[4]{q}}{(2\pi)^4}\sum_{i,j,\calJ,\calP}X^T(-q)\cdot
a_{ij}(\calJ^\calP)\,P_{ij}(\calJ^\calP)\cdot X(q) ~.
}
The $a_{ij}(\calJ^\calP)$ are matrices of kinetic coefficients, called the coefficient matrices, carrying all the information about the propagation and mixing between different degrees of freedom in the same spin-parity sector. By using the orthonormality condition~\eqref{Eq. orthonormality}, it can be checked that the coefficient matrices are related to the kinetic operator by
\bae{\label{Eq. coefficient matrices formula1}
    a_{ij}(\calJ^\calP) =\frac{1}{d(\calJ^\calP)}P_{ij}(\calJ^\calP){}_A{}^B \calO {}^A{}_B ~,
}
where again $d(\calJ^\calP)$ denotes the dimension of the representation coming from the trace of the spin projectors. Note that invariance under diffeomorphism of the metric lowers by one the rank of the coefficient matrices $a(1^-)$ and $a(0^+)$. When there are additional gauge symmetries, the rank is further lowered. One can fix the gauge redundancies by simply suppressing these rows and columns and the remaining invertible submatrices are called $b_{ij}(\calJ^\calP)$.

Dynamical properties of fields are encoded in the saturated propagator, which is given by
\bae{\label{Eq. Saturated propagator}
    \Pi= -\frac12\sum_{i,j,\calJ,\calP}J^T(-q)\cdot
    b^{-1}_{ij}(\calJ^\calP)~ P_{ij}(\calJ^\calP)\cdot J(q) ~,
}
where $J = (\tau, \sigma)^T$ is a set of external sources. Let $C_{ij}$ be the cofactor matrix associated with the invertible submatrix. Then the inverse submatrix $b_{ij}^{-1}(\calJ^\calP)$ can be written as
\bae{
b^{-1}_{ij}(\calJ^\calP) = \frac{1}{\det [b (\calJ^\calP)]}C_{ij}^T ~
}
where $\det [b (\calJ^\calP)]$ is the determinant of the matrix $b_{ij}(\calJ^\calP)$.
Given that the submatrices are polynomial in $q$, poles for massive degrees of freedom can only come from the determinant. Assuming that all fields have distinct masses (except for the graviton), the determinant of a given spin-parity sector with $s$ propagating fields can be written as
\bae{\label{eq. pole det}
\det [b (\calJ^\calP)] = C \cdot q^2 \left(q^2 + M^2_1\right)\left(q^2 + M^2_2\right)\cdots \left(q^2 + M^2_n \right) \cdots \left(q^2 + M^2_s \right) ~,
}
with a constant $C$. Here we should remark that the first $q^2$ corresponds to a massless particle. It appears in the $2^+$ and $0^+$ sectors in our setup, corresponding to the components of the massless graviton.

Let us finally write down the conditions for the absence of ghosts and tachyons. The tachyon-free condition is trivial from the form of Eq.~\eqref{eq. pole det} as
\bae{
	M^2_n > 0 ~.
}
For the physical states, the saturated propagator~\eqref{Eq. Saturated propagator} has non-vanishing and positive residues. So the absence of ghosts requires
\bae{
\Re \left[\mathop{\Res}_{q^2 \to - M^2_n} ~ (\Pi)\right] > 0 ~.
}
For massive fields, this inequality can be evaluated only with the inverse matrix $b_{ij}^{-1}(\calJ^\calP)$. Taking care of an additional sign from the spin projectors, the massive no-ghost condition becomes~\cite{Lin:2018awc}
\bae{\label{eq. ghost freedom}
\mathop{\Res}_{q^2 \to - M^2_n}\sum_i\left[-1\cdot (-1)^{n_\mathrm{L}^{(i)}}\cdot b_{ii}^{-1} \right] > 0 ~,
}
where $n_\mathrm{L}^{(i)}$ is the number of longitudinal operators carried by the degree of freedom labelled by $i$, which can be counted from the table~\ref{Table irreps}.

\section{Particle spectrum of UV theories}\label{sec. particle spectrum}
We turn to the particle spectrum in the \ac{UV} theories. In the antisymmetric case, we start with the Lagrangian~\eqref{eq. UV lag torsion} without relations among the UV parameters. We then study how the large-$N$ limit changes the spectrum. In the symmetric case, we work on the action with the large-$N$ limit~\eqref{eq. large N nonmetricity}.
\subsection{Antisymmetric metric-affine gravity}\label{sec. Antisymmetric metric-affine gravity}
Recall that the UV theory without the large-$N$ limit in antisymmetric \ac{MAG} is given in Eq.~\eqref{eq. UV lag torsion} by
\beae{\label{eq. UV lag torsion re}
    \calL_{\mathrm{UV}}^{\mathrm{T}} =
    \bmte{\frac{\Mpl^2}{2}R + \alpha R^2 -\frac{1}{2}b^{RT}_5 R \trdivT1 -\frac{1}{2}b^{TT}_9 (\trdivT1)^2 
    \\
    -\frac{\Mpl^2}{2}\left(c_1^{TT} M^{TT}_1 + c_2^{TT} M^{TT}_2 + c_3^{TT} M^{TT}_3 \right) ~.}
}
Here, we rename the mass parameters $m_i^{TT}$ as $c^{TT}_i$ for a unified notation. 
With use of the spin projector formalism explained in Sec.~\ref{sec SPO formalism}, it is possible to extract the kinetic structure of the theory in a systematic way. Besides the spin-2 graviton, the \ac{UV} theory~\eqref{eq. UV lag torsion re} has dynamical components only in the $0^+$ sector. The nondegenerate coefficient matrix in the $0^+$ sector is given by
\bae{\label{Eq. coefficient matrix 0p}
	b(0^+) = \mqty(-\frac{1}{2}\tilde{c}\Mpl^2-\frac{3}{2}b^{TT}_9 q^2 & i\frac{3b^{RT}_5}{\sqrt{2}\Mpl} q^3 \\ -i\frac{3b^{RT}_5}{\sqrt{2}\Mpl} q^3  & 2 q^2 +\frac{24 \alpha}{\Mpl^2}q^4 ) ~,
}
where $\tilde{c}$ is defined as a combination of the UV parameters $c^{TT}$ by
\bae{
	\tilde{c} \coloneqq 2 c^{TT}_1 + c^{TT}_2+ 3 c^{TT}_3 ~.
}
The determinant of the matrix~\eqref{Eq. coefficient matrix 0p} is simply given by
\bae{\label{eq. det b torsion}
\det [b (0^+)] = -\tilde{c}\Mpl^2 q^2 -3 \left(b^{TT}_9+4\tilde{c}\alpha\right)q^4 -\frac{9}{2}\left\{(b^{RT}_5)^2 + 8 b^{TT}_9 \alpha\right\}\frac{q^6}{\Mpl^2} ~,
}
implying that there can be two massive degrees of freedom in the spectrum.
\subsubsection[Without large-$N$ limit]{\boldmath Without large-$N$ limit}
\begin{figure}
    \centering
    \includegraphics[width=\hsize]{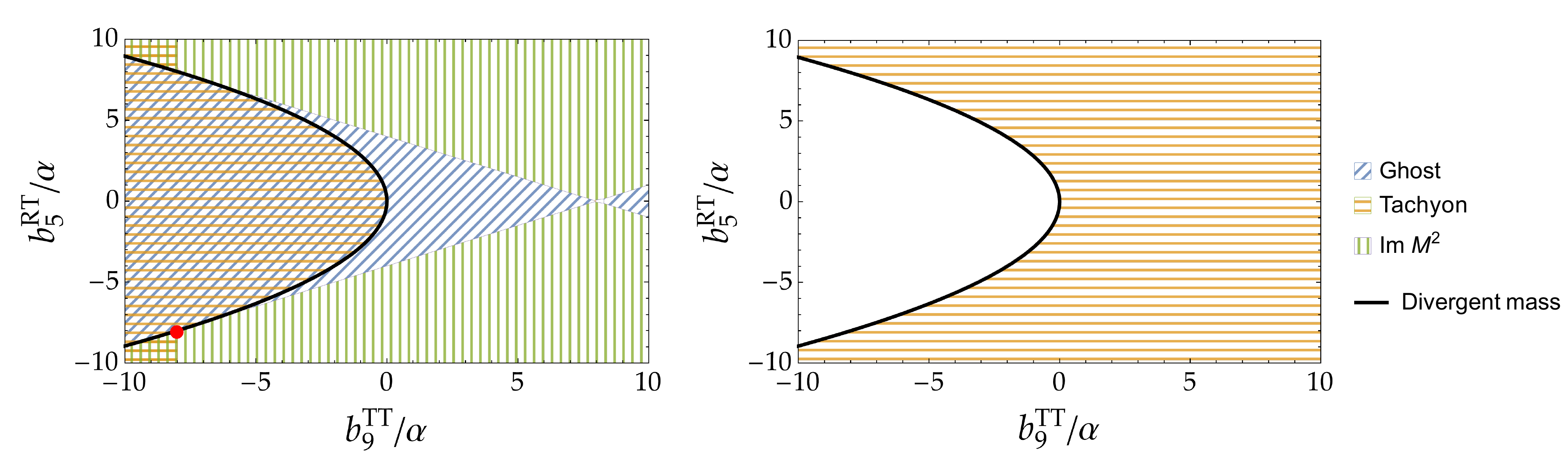}
    \caption{A parameter space with either ghosts or tachyons for two different values of $\tilde{c}$ (\emph{Left}:~$\tilde{c}=2$ and \emph{Right}:~$\tilde{c}=-2$). Hatched regions are prohibited as the theory contains pathological particles. The black line denotes the parameter region where the number of propagating particles is reduced. The red point ($\tilde{c}=2$ and $b^{TT}_9/\alpha = b^{RT}_5/\alpha =-8$) corresponds to the Palatini-Higgs inflation. If we take $\tilde{c}=2$, two particles can be either ghost or tachyon, or have imaginary squared masses in all the plausible regions. With a negative $\tilde{c}$, there exists a healthy region in white.
    }
    \label{Fig. Torsion}
\end{figure}

Let us now study in which parameter regions the \ac{UV} theory~\eqref{eq. UV lag torsion re} is free of pathological propagation. By rewriting the determinant~\eqref{eq. det b torsion} as
\bae{
\det [b (0^+)] \propto q^2 \left(q^2+M_{+}^2\right)\left(q^2+M_{-}^2\right) ~,
}
we find that the masses of the two propagating particles are 
\beae{\label{Eq. Torsion mass}
	M_{\pm}^2 = \frac{b^{TT}_9 + 4 \tilde{c} \alpha \pm \sqrt{-2 \tilde{c} \left( b^{RT}_5 \right)^2 + \left(b^{TT}_9 - 4 \tilde{c} \alpha \right)^2}}{3\left\{\left( b^{RT}_5 \right)^2 + 8 b^{TT}_9 \alpha\right\}} \Mpl^2 ~.
}
To have real and non-tachyonic masses, we have to require
\bae{
-2 \tilde{c} \left( b^{RT}_5 \right)^2 + \left(b^{TT}_9 - 4 \tilde{c} \alpha \right)^2 > 0 ~, \quad M_{\pm}^2 > 0 ~.
}
The propagating particles are ghost-free if the inequality~\eqref{eq. ghost freedom} is satisfied, where two conditions in this case are
\bae{
-\frac{\left( b^{RT}_5 \right)^2 \left(8+ 3 b^{TT}_9 \pm 3f - 12 \tilde{c} \alpha \right) +8\alpha \left\{b^{TT}_9\left(4+3b^{TT}_9 \pm f - 12 \tilde{c} \alpha \right) - 4\left(\pm f + 4 \tilde{c} \alpha\right) \right\}}{\pm 12 f \left\{\left( b^{RT}_5 \right)^2 + 8 b^{TT}_9 \alpha\right\}}> 0 ~,
}
where we define
\bae{
f \coloneqq \sqrt{-2 \tilde{c} \left( b^{RT}_5 \right)^2 + \left(b^{TT}_9 - 4 \tilde{c} \alpha \right)^2} ~.
}
Fig.~\ref{Fig. Torsion} shows the parameter region where the particles propagate pathologically. The left figure is depicted with $\alpha=10^5$ and $\tilde{c}=2$ (recall that the Palatini-Higgs inflation corresponds to $\tilde{c}=2$). It shows that the particles become either ghosts or tachyons in a plausible parameter space. We confirm that this holds true if we take different values of $\tilde{c}$, e.g., $\tilde{c}=2$, $4$, $6$, $8$, and $10$, implying that a positive value of $\tilde{c}$ is not allowed theoretically. The right figure corresponds to a choice $\alpha=10^5$ and $\tilde{c}=-2$, which shows that a certain range of parameters produces a healthy spectrum in the \ac{UV} theory. In both figures, on the black line, the number of propagating particles is reduced. This region is not theoretically prohibited but a separate analysis needs to be performed.
\subsubsection[With large-$N$ limit]{\boldmath With large-$N$ limit}
The number of propagating degrees can be reduced if the \ac{UV} parameters satisfy
\bae{
(b^{RT}_5)^2 + 8 b^{TT}_9 \alpha = 0 ~,
}
which removes the $q^6$ term in the determinant~\eqref{eq. det b torsion}. The large-$N$ limit of the deformed theory is in this category, which we investigate in this section.

The large-$N$ discussion motivates us to restrict the \ac{UV} theory as Eq.~\eqref{eq. UV lag largeN}. There, the terms up to the quadratic order are summarized as
\bae{\label{eq. UV lag torsion Large N}
\calL_{\mathrm{UV}, N}^{\mathrm{T}} \supset \frac{\Mpl^2}{2}R + \alpha \left(R + x \trdivT1 \right)^2 -\frac{\Mpl^2}{2}\left(c_1^{TT} M^{TT}_1 + c_2^{TT} M^{TT}_2 + c_3^{TT} M^{TT}_3 \right) ~,
}
where $x$ is a parameter of order one which is related to the UV parameters as
\bae{
b^{RT}_5 =-4\alpha x ~, \quad b^{RT}_9 = -2\alpha x^2 ~.
}
Note that $x=2$ corresponds to the Palatini-Higgs inflation. By means of the spin projector formalism, it is easy to see that, except for the massless graviton stored in the spin-$2$ sector, the Lagrangian~\eqref{eq. UV lag torsion Large N} has only one dynamical degree of freedom in the $0^+$ sector. The determinant of the nondegenerate coefficient matrix for the $0^+$ sector reads
\bae{\label{Eq. determinant 0p large-N}
	\det [b(0^+)] = - \tilde{c}\Mpl^2 q^2 
	-6\alpha\left(2\tilde{c}-x^2\right)q^4 ~.
	}
As expected, the Palatini-Higgs limit $\pqty{\tilde{c},x}=\pqty{2,2}$ eliminates the $q^4$ term, indicating the absence of a massive particle.
By writing the determinant~\eqref{Eq. determinant 0p large-N} as
\bae{
\det [b(0^+)] \propto q^2 \left(q^2 + M^2\right) ~,
}
we can obtain the mass of the propagating field as
\bae{\label{Eq. anti mass large-N}
	M^2 =\frac{\tilde{c}\Mpl^2}{6\alpha (2\tilde{c} -x^2)} ~.
}
If we take a limit $x=0$, vanishing torsion, the mass becomes
\bae{
M^2 =\frac{\Mpl^2}{12\alpha} ~,
}
which is nothing but the scalaron mass coming from the $R^2$ operator.

\begin{figure}
    \centering
    \includegraphics[width=0.75\hsize]{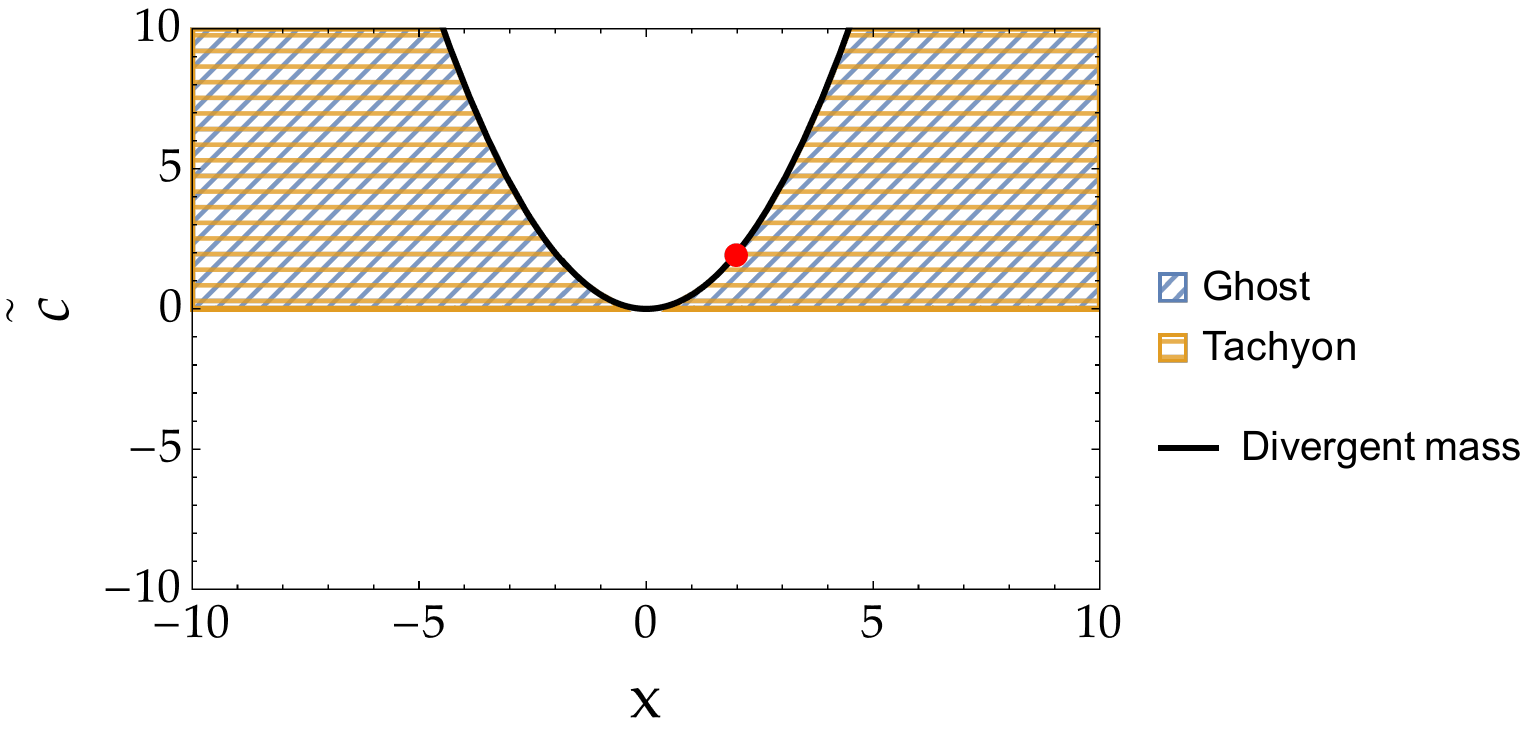}
    \caption{A parameter space with either ghosts or tachyons. Hatched regions are prohibited as the theory contains a pathological particle. On the black line, the propagating particle becomes infinitely heavy, indicating that no propagating scalar appears in the UV regime. The red point corresponds to the Palatini-Higgs inflation. If we take a positive $\tilde{c}$, the remaining parameter $x$ has to be within a particular region to avoid the pathological propagation. With a negative $\tilde{c}$, the propagating particle is free of these pathologies.} 
    \label{Fig. LargeN-T}
\end{figure}

We turn to the ghost- and tachyon-free conditions keeping $\tilde{c}$ unfixed. Assuming $\alpha$ is positive, the tachyon-free condition is trivially obtained as 
\bae{
\frac{\tilde{c}}{2\tilde{c}-x^2}>0 ~,
}
and, from Eq.~\eqref{eq. ghost freedom}, the ghost freedom requires
\bae{
- \frac{-6\tilde{c}^2\alpha + x^2(-1+3\tilde{c}\alpha)}{3(x^2-2\tilde{c})^2\alpha} > 0 ~,
}
which is equivalent to 
\bae{
6\tilde{c}^2\alpha + x^2(1 - 3\tilde{c}\alpha) > 0 ~.
}
Fig.~\ref{Fig. LargeN-T} shows a region of pathological propagation by taking $x$ and $\tilde{c}$ as free parameters, while fixing $\alpha$ to be $10^5$. It is clear from the figure that a region without pathology exists with both positive and negative values of $\tilde{c}$. While $x$ can be arbitrary with a negative value of $\tilde{c}$, a positive value of $\tilde{c}$ requires that $x$ is constrained within a certain range. It is interesting to observe that the Palatini-Higgs inflation is located on the black line on which the number of propagating particles is further reduced. With a set of parameters on this black line, the large-$N$ limit is no longer a successful approach to find out weakly coupled UV completion.
\subsection{Symmetric metric-affine gravity}\label{sec. Symmetric metric-affine gravity}
Let us move on to the case of symmetric \ac{MAG}. Recall that the \ac{UV} theory is defined by Eq.~\eqref{eq. UV lag nonmetricity}:
\beae{\label{eq. UV lag nonmetricity again}
    \calL_{\mathrm{UV}}^{\mathrm{Q}} \coloneqq 
    \bmte{\frac{\Mpl^2}{2}R + \alpha R^2 -\frac{1}{2}b^{RQ}_6 R \trdivQ1 -\frac{1}{2}b^{RQ}_7 R \trdivQ2
    -\frac{1}{2}b^{QQ}_{14} (\trdivQ1)^2 
    \\ 
    -\frac{1}{2}b^{QQ}_{15} (\trdivQ2)^2 -\frac{1}{2}b^{QQ}_{16} \trdivQ1\cdot \trdivQ2
    -\sum_{i}\frac{\Mpl^2}{2}c_i^{QQ} M^{QQ}_i ~, }
}
with $(i=1,2,3,5)$. Without any relations among the UV parameters, it is easy to check from the coefficient matrix that there exist three massive particles in the $0^+$ sector. As a full classification of ghost- and tachyon-free \ac{MAG} in the theory~\eqref{eq. UV lag nonmetricity again} is a daunting task, we will only consider the large-$N$ limit of the deformed theory~\eqref{eq. large N nonmetricity}.

In the large-$N$ limit, the relevant terms in the UV theory are given by
\beae{
    \calL_{\mathrm{UV},N}^{Q}
    \supset\bmte{\frac{\Mpl^2}{2}R+ \alpha \left(R + x\trdivQ1 + y\trdivQ2 \right)^2
    \\
    -\frac{\Mpl^2}{2}\left(-\frac{1}{4} M^{QQ}_1 + \frac{1}{2} M^{QQ}_2 + \frac{1}{4} M^{QQ}_3 -\frac{1}{2} M^{QQ}_5 \right) ~.}
}
It is easy to see that the vanilla Palatini-Higgs inflation with leading order quantum corrections in the large-$N$ limit is recovered by taking $\pqty{x,y}=\pqty{-1,1}$. Interestingly, as in the case of torsion, the \ac{UV} theory in the large-$N$ limit contains only one dynamical degree of freedom in the $0^+$ sector on top of the graviton. The determinant of the nondegenerate coefficient matrix is given by
\bae{\label{Eq. Qdeterminant 0plarge-N}
	\det [b(0^+)] = \frac{3}{32}\Mpl^6 q^2 
	+\frac{\alpha}{8}\left(9-4x^2+16 xy + 11y^2\right) \Mpl^4 q^4 ~,
	}
implying that the mass of a propagating scalar field is
\bae{\label{Eq. mass 0plarge-N}
M^2 = \frac{3}{4\left(9-4x^2 + 16 xy + 11y^2\right) \alpha}\Mpl^2 ~.
}

\begin{figure}
    \centering
    \includegraphics[width=0.75\hsize]{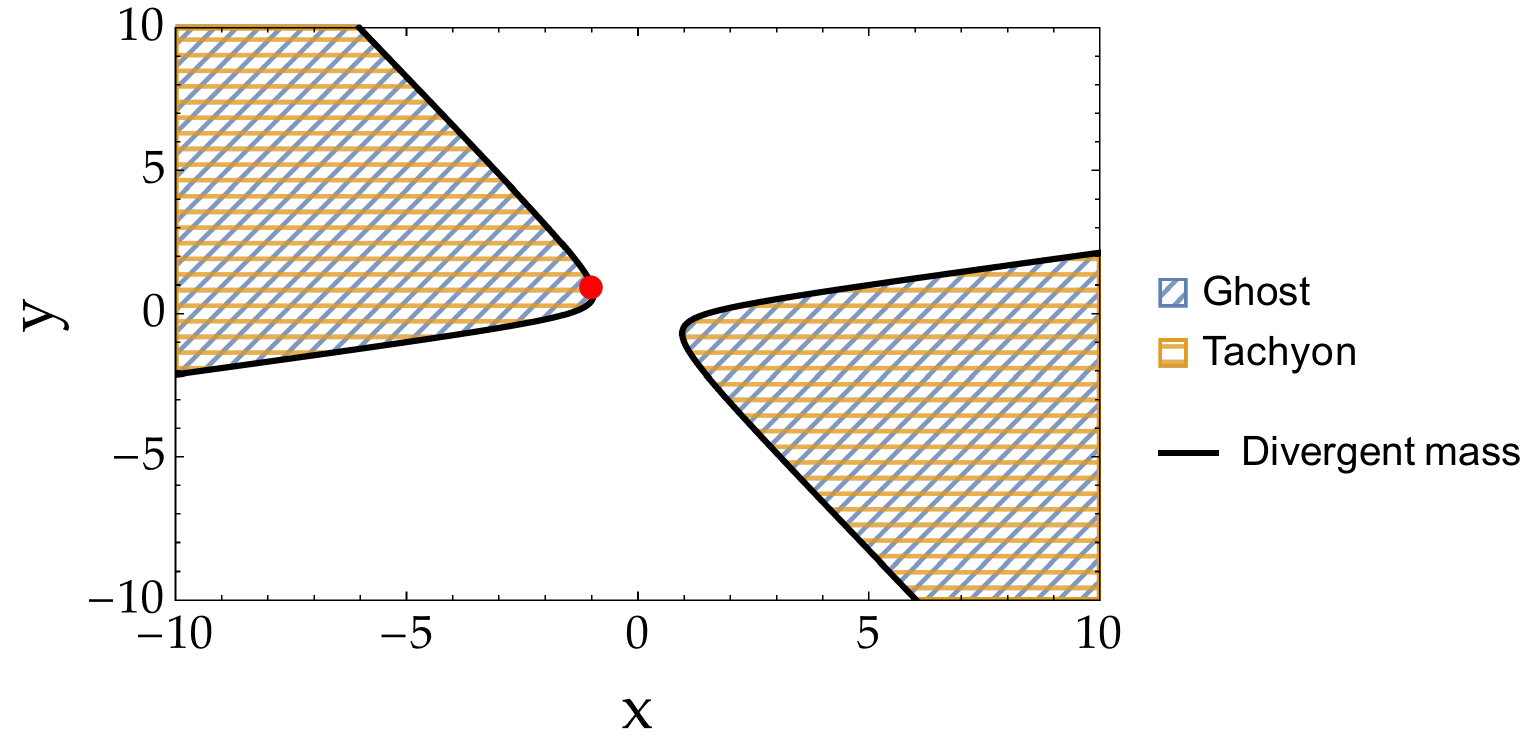}
    \caption{A region with either ghost or tachyon with $\alpha =10^5$. The hatched region is prohibited due to the existence of a pathological particle, while the white region is free of the pathology. On the black solid line, a UV particle becomes infinitely heavy, so that it disappears from the spectrum. The red point corresponds to the Palatini-Higgs inflation.} 
    \label{Fig. pathology Q}
\end{figure}

We are now ready to search for the parameter space without pathological propagation. The absence of a tachyon is simply achieved by requiring $M^2>0$. The absence of a ghost requires that the following inequality is satisfied:
\bae{
-\frac{x^2(40-36\alpha)+81\alpha+8xy(7+18\alpha)+y^2(52+99\alpha)}{2\left(9-4x^2 + 16 xy + 11y^2\right)^2 \alpha} > 0 ~.
}
Fig.~\ref{Fig. pathology Q} shows in what parameter region the theory is free of a pathological propagation. The hatched region is prohibited to have a healthy UV theory in the large-$N$ limit. On the black solid line where the denominator of Eq.~\eqref{Eq. mass 0plarge-N} becomes zero, the mass of the propagating field is divergent. As it shows that there is no propagating particle in the UV regime, again with parameters on this black line, the large-$N$ limit cannot find out weakly coupled UV completion.
\section{Conclusions}\label{sec. Conclusions}
In this paper, we have introduced a new class of Higgs inflation in the framework of parity-even \ac{MAG} by deforming the Palatini-Higgs inflation to its neighbor. By mainly relying on the large-$N$ limit, we have constructed a corresponding UV theory and investigated the particle spectrum around the Minkowski space. 
We have found that the absence of ghosts and tachyons can put constraints on parameters in the deformed low-energy theory and showed explicitly in what parameter region UV particles are free of these pathologies.  In Sec.~\ref{sec. Antisymmetric metric-affine gravity}, we have focused on the case where the connection is antisymmetric. The left panel of Fig.~\ref{Fig. Torsion} shows the region of pathological propagation with a positive $\tilde{c}$ and $\alpha$ without relying on the large-$N$ limit. In this case, we have two massive propagating particles and have found that they become either ghosts or tachyons or have imaginary square masses in a plausible parameter space, so that a positive $\tilde{c}$ with two massive UV particles is not theoretically allowed. On the other hand, if we take a negative value of $\tilde{c}$, there exists the parameter space in which pathological propagation is safely avoided. Once we rely on the large-$N$ limit, the number of propagating fields is reduced and both positive and negative $\tilde{c}$ have the healthy parameter space while another parameter $x$ is subjected to a constraint with a positive value of $\tilde{c}$ as can be seen in Fig.~\ref{Fig. LargeN-T}.
In Sec.~\ref{sec. Symmetric metric-affine gravity}, we perform the same analysis in symmetric \ac{MAG}, but in the large-$N$ limit only. There is only one propagating UV particle and the healthy parameter region can be read out from Fig.~\ref{Fig. pathology Q}. 

This work can be placed as a first step of a full classification of low-energy theories from the viewpoint of the theoretical consistency, while it cannot be rigorous since we rely on the large-$N$ limit. There exist several directions to extend this work. First, it should be checked whether or not inflation can take place with a set of parameters in this paper with which pathological propagation is avoided and a new cutoff scale is properly uplifted to the Planck scale. Once these two are satisfied, the model will be the first Higgs inflation in \ac{MAG} that can be completed up to the Planck scale.
Second, it is possible to broaden the space of Higgs inflation models since MAG contains other operators we do not deal with here. It might be important and interesting to check their inflationary phenomenology exhaustively. 

\section*{Acknowledgments}
We would like to thank Sebastian Zell for the fruitful discussions about the small deformation of the Palatini-Higgs inflation which inspired this work. We are also grateful to Roberto Percacci for useful discussions. This work is supported by JSPS KAKENHI Grants 
No.~JP22KJ1600 (Y.M.) and No.~JP24K07047 (Y.T.). 
We utilize the free software packages \texttt{xAct}, \texttt{xTensor}, \texttt{xPert}, and \texttt{xTras}.
Feynman diagrams are drawn by \texttt{TikZ-Feynhand}~\cite{Dohse:2018vqo} and \texttt{TikZ-Feynman}~\cite{Ellis:2016jkw}.

\bibliography{Bib}
\bibliographystyle{utphys}
\end{document}